\documentclass[preprint,pra,superscriptaddress,amsmath,amssymb]{revtex4-1}
\usepackage{amsmath}
\usepackage{amsfonts}
\usepackage{amssymb}
\usepackage{graphicx}
\usepackage{color}
\usepackage{txfonts}
\usepackage[titletoc]{appendix}
\usepackage[colorlinks={true}]{hyperref}
\hypersetup{citecolor={blue}, filecolor={blue}, linkcolor={blue}, urlcolor={blue}}
\begin{document}
\preprint{}
\title{Orientational quantum revivals induced by a single-cycle terahertz pulse}
\author{Chuan-Cun Shu}
\affiliation{Hunan Key Laboratory of Super-Microstructure and Ultrafast Process, School of Physics and Electronics, Central South University,
Changsha 410083, China}
\author{Qian-Qian Hong}
\affiliation{Hunan Key Laboratory of Super-Microstructure and Ultrafast Process, School of Physics and Electronics, Central South University,
Changsha 410083, China}
\author{Yu Guo}
\email{guoyu@csust.edu.cn}
\affiliation{Hunan Provincial Key Laboratory of Flexible Electronic Materials Genome Engineering, School
of Physics and Electronic Science, Changsha University of Science
and Technology, Changsha 410114, China}
\affiliation{Key Laboratory of Low Dimensional Quantum Structures and Quantum Control
(Hunan Normal University), Ministry of Education, Changsha 410081, China}
\author{Niels E. Henriksen}
\email{neh@kemi.dtu.dk}
\affiliation{Department of Chemistry, Technical University of Denmark, Building 207, DK-2800 Kgs. Lyngby, Denmark}
 \begin{abstract}
The phenomenon of quantum revivals resulting from the self-interference of wave packets has been observed in several quantum systems and utilized widely in spectroscopic applications. Here, we present a combined analytical and numerical study on the generation of orientational quantum revivals (OQRs) exclusively using a  single-cycle THz pulse. As a proof of principle, we examine the scheme in the linear polar molecule HCN with experimentally accessible pulse parameters and obtain strong field-free OQR without requiring the condition of the sudden-impact limit.  To visualize the involved quantum mechanism,  we derive a three-state model using the Magnus expansion of the time-evolution operator.  Interestingly,  the THz pulse interaction with the electric-dipole moment can activate direct  multiphoton processes, leading to OQR enhancements beyond that induced by a rotational ladder-climbing mechanism from the rotational ground state. This work provides an explicit and feasible approach toward quantum control of molecular rotation, which is at the core of current research endeavors with potential applications in atomic and molecular physics, photochemistry, and quantum information science.
\end{abstract} 
\maketitle

\begin{figure}[!t]\centering
\resizebox{0.45\textwidth}{!}{%
  \includegraphics{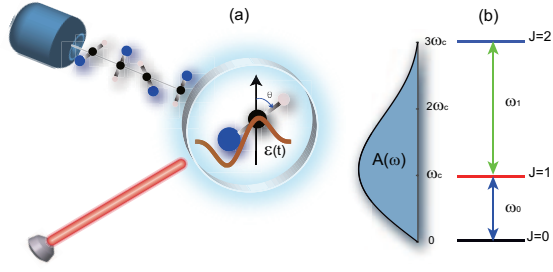}
} \caption{Schematic of quantum control of a linear polar molecule with a single-cycle THz pulse. (a) A linearly polarized THz pulse $\mathcal{E}(t)$ interacts with gas-phase  HCN molecules, where $\theta$ denotes the angle between the rotor axis and the pulse polarization.  (b) A three-state model consists of rotational states $J=0$, 1 and 2, which fall within the frequency distribution $A(\omega)$ of the pulse centered at $\omega_c$. $\omega_0$ and $\omega_1$  correspond to the transition frequencies between rotational states. } \label{fig1}
\end{figure}

Quantum revival (QR), i.e., a periodic recurrence of wave packets, is a fundamental time-dependent interference phenomenon for states with quantized energies \cite{QR1}.  This phenomenon closely connects to quantum echoes \cite{QE}, quantum Talbot effect \cite{QTE}, quantum scars \cite{QS}, and molecular charge migration \cite{mcm,mcm1}, and therefore is of broad interest in physics, chemistry, and information science. QRs have been observed in semiconductor wells \cite{qric}, ion traps \cite{qri}, and graphene \cite{qrg}. QR can also appear in molecules by creating
a rotational wave packet (i.e., a coherent superposition of rotational states), leading to time-dependent aligned or oriented molecules \cite{qr-TS,IRPC2010,IJC2012,qr-sugny}.\\ \indent
Despite various schemes proposed to generate the rotational wave packets \cite{qr-TS,IRPC2010,IJC2012,qr-sugny,qr-misha,Juan,qr-niels}, the unique properties of terahertz radiation with a frequency range between 0.1 and 1 THz and simultaneously with high peak fields offer excellent opportunities to control rotational motions of molecules \cite{THz}. An intense THz pulse can force the molecular dipoles to transiently orient along the polarization axis of the optical field, giving rise to  orientational quantum revivals (OQRs)-a phenomenon known as field-free molecular orientation \cite{qr-niels2,qr-sakai,qr-mjj,qr-kling,qr-wj}. The finding of this fascinating phenomenon opens a new avenue to rotating the molecular sample toward the desired direction in the lab frame  and  has potential applications for studying the orientation dependence of photon-molecule and molecular interactions \cite{jacs2009,wj2,science2013,nm,HHG-LU,PT1}.\\ \indent
Following original proposals \cite{niels3,qr-niels2,dion}, a great effort was put to realize OQRs in the sudden-impact limit by using half-cycle THz pulses \cite{qr-sugny,pra2006,sugny2,shu2}, which feature a large asymmetry in the magnitude of the positive and negative peak values. Since the effect of
the long weak negative tail on excitations can be neglected, the short central part with a non-zero (time-integrated) area transfers impulsively an angular momentum to the molecule. It creates the rotational wave packet, leading to the ``kick" mechanism of OQRs. Recently, this OQR phenomenon was carried forward in the sudden-impact
limit by using a single THz pulse with a zero time-integrated area \cite{Sugny4,shu3,shuJCP,ex0,ex00,J2, Sugny6,PRL2020}, which generates the rotational wave packet based on a resonant-excitation mechanism.
To assist more rotational states to be excited by the single-cycle THz pulse, a hybrid scheme \cite{jiro,shu4} that has been examined in experiments \cite{ex1,ex2} applies an intense nonresonant ultrashort pulse to align the molecules prior to the THz irradiation,  leading to a substantial enhancement of the degree of orientation. However, it remains a challenging task to obtain strong OQR  by using exclusively a single THz pulse, and a fundamentally important but largely
unexplored question is whether the underlying physics has to require the condition of the sudden-impact limit.\\ \indent
In this work, we present a theoretical study to show a large OQR by using an experimentally accessible single-cycle THz pulse with a zero-area and a comparable duration to the rotational period of molecules. We derive a theoretical model  to reveal the underlying physics via the Magnus expansion of the time-evolution operator.  Interestingly, we find that the interaction of  the THz pulse  with the electric-dipole moment (EDM) can activate direct multiphoton processes via higher-order Magnus terms, enhancing the OQR amplitude over the level governed by the first-order Magnus term. This work  provides an explicit model for generating OQRs without the use of the ``kick" mechanism and  a way to visualize  multiphoton processes induced by strong THz fields. \\ \indent
The general concept of our scheme is illustrated in Fig.  \ref{fig1} for generating OQRs by using a single-cycle THz pulse. We consider the linear polar molecule HCN in its ground vibronic state described as  a rigid rotor with the rotational constant  $B$ (1.457 cm$^{-1}$) and the EDM $\mu$ (2.89 Debye). The molecule is driven by a linearly polarized single-cycle THz pulse $\mathcal{E}(t)=\mathcal{E}_0\sin^2\left(\pi t/T\right)\cos\left(\omega_ct+\phi_c\right)
$ with a peak field strength $\mathcal{E}_0$, duration $T$,  central frequency $\omega_c$, and absolute phase $\phi_c$ \cite{Sugny4}.  It turns on  at $t=0$ and off at $t=T$ with a duration $T=2\pi/\omega_c$ (i.e., one optical cycle of the pulse).  The use of the phase $\phi_c=\pi/2$ can exclude the  DC component in its frequency spectrum, i.e., by satisfying a zero-area $\int_0^{T}dt\mathcal{E}(t)=0$.
 The molecular Hamiltonian reads $\hat{H}(t)=\hat{H_0}+\hat{V}(t)$ with  the field-free Hamiltonian $\hat{H}_0=B\hat{L}^2$  and the time-dependent interaction potential $\hat{V}(t)=-\mu \mathcal{E}(t)\cos\theta$, where $\hat{L}$ is an angular momentum operator and $\theta$ denotes the angle between the rotor axis and the pulse polarization. \\ \indent
 We utilize $\hat{V}(t)$ to generate a superposition of rotational eigenstates $|JM\rangle$ with quantum numbers $J$ and $M$.  For a linearly polarized excitation, the quantum number $M$ associated with the projection of the angular momentum along the polarization
axis is conserved, and therefore the time-dependent wave function of the molecule reads ($\hbar=1$)
 \begin{eqnarray}\label{wf}
|\psi_{J_0M}(t)\rangle=\sum_{J'=0}c_{J'M}(t)e^{-iE_{J'}t}|J'M\rangle
\end{eqnarray}
where  $|JM\rangle$ satisfy $\hat{L}^2|J'M\rangle=E_{J'}|JM\rangle$ with eigenenergies  $E_{J'}=BJ'(J'+1)$, and $c_{J'M}$ are  the expansion coefficients of $|J'M\rangle$. We use a unitary operator  $\hat{U}(t,t_0)$ to describe  the time evolution of the system  from the initial time $t_0$ to a given time $t$, which has a solution
\begin{eqnarray}
\hat{U}(t, t_0)=\hat{U}(t_0,t_0)-i\int_{t_0}^tdt'\hat{H}_I(t)\hat{U}(t',t_0)
\end{eqnarray}
where $\hat{U}(t_0,t_0)=\mathbb{I}$ and $\hat{H}_I(t)=\exp(i\hat{H}_0t)[-\hat{\mu}E(t)]\exp(-i\hat{H}_0t)$ with $\mu_{JJ'}=\mu\langle J'M|\cos\theta|JM\rangle$ as the matrix elements of the dipole operator $\hat{\mu}$.  The coefficients $c_{J'M}(t)$ in Eq. (\ref{wf}) can be calculated by $c_{J'M}(t)=\langle J'M|\exp(iE_{J'}t)\hat{U}(t,t_0)|J_0M\rangle$ starting from $|J_0M\rangle$.\\ \indent
The thermally averaged expectation value of $\cos\theta$ (i.e., the degree of orientation) can be given by
\begin{eqnarray} \label{cos}
\left\langle\cos\theta\right\rangle(t)&=&\sum_{J_0=0}^{\infty}\sum_{M=-J_0}^{J_0}\mathcal{P}(J_0)\sum_{J=0}^{\infty}2\left|c_{J+1M}\left(t\right)\right|\left|c_{JM}\left(t\right)\right|\\ \nonumber
&&\times\mathcal{M}_{J+1,J}\cos\left(\omega_Jt-\phi_J\right)
\end{eqnarray}
 where $\mathcal{P}(J_0)$ is the Boltzmann distribution associated with the initial  states $J_0$,  the transition matrix $\mathcal{M}_{J+1,J}=\langle J+1M|\cos\theta|JM\rangle=\sqrt{(J+1)^2-M^2}/\sqrt{(2J+1)(2J+3)}$, rotational frequencies $\omega_J=E_{J+1}-E_{J}=2(J+1)B$, and the relative phases $\phi_J=\arg(c_{J+1M}(t))-\arg(c_{JM}(t))$.  All frequencies are equal to an integer times $2B$, and therefore OQRs will occur  at a time interval $\tau=\pi/B$ by generating the coherent superposition of rotational states as defined by Eq. (\ref{wf}).   \\ \indent
 \begin{figure}[!t]\centering
\resizebox{0.48\textwidth}{!}{%
  \includegraphics{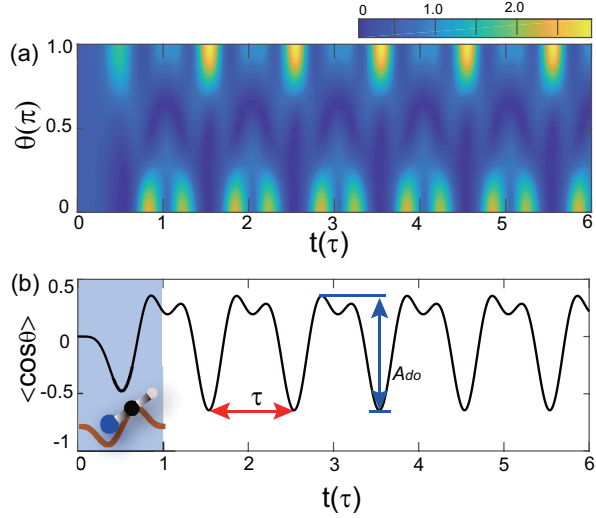}
} \caption{Orientational quantum revivals (OQR) by a single-cycle THz pulse. (a) Evolution of the  wave packets for the molecular ensemble at low temperature 2 K, (b) the corresponding degree of orientation as a function of time. The red double arrow shows the revival time $\tau$ (11.45 ps for HCN), and the blue double arrow denotes the OQR amplitude.} \label{fig2}
\end{figure}
To demonstrate the OQRs by Eq. (\ref{cos}), we consider a  molecular sample  at a low temperature $T=2$ K, for which initial states $J_0=0$ and $J_0=1$ ($M=-1, 0, 1$) make significant contributions to the ensemble. Experimentally it has been possible to generate an intense single-cycle pulse with a record peak strength  up to $3.0\times10^7$ V/m and a central frequency around 0.1 THz \cite{SCTP}.  We first perform a simulation at resonant excitation with  $\omega_c=\omega_0$ (0.09 THz, the corresponding pulse duration is equal to the rotational period) and $\mathcal{E}_0=7.0\times10^6$ V/m far below the experimental limit. Figure \ref{fig2} shows the  time-dependent probability density $|\psi(\theta, t)|^2$ and the corresponding degree of orientation $\langle\cos\theta\rangle(t)$. We can see a periodic recurrence of wave packets in  Fig. \ref{fig2}(a), showing asymmetric angular distributions with respect to the polarization axis of the field. Equally spaced OQRs emerge with a revival time $\tau$ (i.e. $2\pi/
\omega_0=11.45$ps) as described by Eq. (\ref{cos}).  The degree of orientation has a local maximum of  $\langle\cos\theta\rangle_{max}=0.36$ and a local minimum  of $\langle\cos\theta\rangle_{min}=-0.64$. We define their difference as a new parameter to describe the OQR amplitude, i.e., $A_{OQR}=\langle\cos\theta\rangle_{max}-\langle\cos\theta\rangle_{min}$, varying  in the range [0, 2]. \\ \indent
\begin{figure}[!t]\centering
\resizebox{0.48\textwidth}{!}{%
  \includegraphics{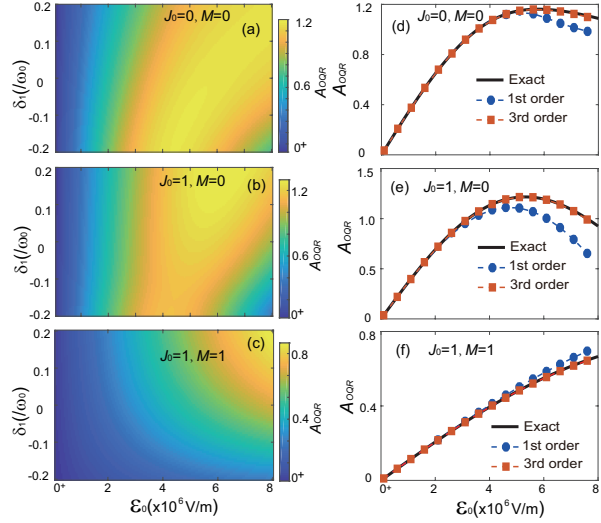}
} \caption{The dependence of the OQR amplitude $A_{OQR}$ on the laser parameters. (a)-(c)  The  amplitude $A_{OQR}$ versus the field strength $\mathcal{E}_0$ and the deturning $\delta_1=\omega_c-\omega_0$ for the molecule initially in $|00\rangle$, $|10\rangle$ and $|11\rangle$,  respectively. (d)-(f) The comparisons  of the exactly calculated  $A_{OQR}$ (black line) at $\delta_1=0$ versus $\mathcal{E}_0$ with the first-order (blue line and circles)  and third-order (orange line and circles) Magnus descriptions. } \label{fig3}
\end{figure}
To access the underlying OQR mechanism induced by the single-cycle THz pulse,  we  perform  the Magnus expansion on the unitary time-evolution operator \cite{pr:470:151,pra:92:063815}   \begin{eqnarray} \label{UO}
\hat{U}(t, t_0)=\exp\left[\sum_{n=1}\hat{S}^{(n)}(t)\right],
\end{eqnarray}
where the first three leading terms can be given by means of the Baker-Campbell-Hausdorff formula as $\hat{S}^{(1)}(t)=-i\int_{t_0}^tdt_1\hat{H}_I(t_1)$, $\hat{S}^{(2)}(t)=(-i)^2/2\int_0^t dt_1\int_0^{t_1}dt_2[\hat{H}_I(t_1),\hat{H}_I(t_2)]$, and $\hat{S}^{(3)}(t)=(-i)^3/6\int_0^t dt_1\int_0^{t_1}dt_2\int_0^{t_2}dt_3[\hat{H}_I(t_1),[\hat{H}_I(t_2), \hat{H}_I(t_3)]]$. As shown in Fig. \ref{fig1} (b), the energy differences between neighboring rotational states of $J\leq2$ are comparable to the frequency components of the THz pulse. Thus, we restrict our analysis within a three-state model consisting of rotational states $J=0, 1$, and 2.  By expanding $\hat{U}(t, t_0)$ to  the first-order Magnus term, the unitary operator  can be described by \cite{shupra,shuprl}
\begin{eqnarray} \label{1stU}
\hat{U}^{(1)}(t,t_0)&=&\sum_{p=-,0, +}\exp(i\lambda_{p}(t))|\lambda_{p}\rangle\langle\lambda_{p}|
\end{eqnarray}
where  $\lambda_0(t)=0$ and $\lambda_{\pm}(t)=\pm\beta(t)=\sqrt{|\beta_{0}|^2+|\beta_{1}(t)|^2}$ are the eigenvalues of $-i\hat{S}^{(1)}(t)$,  and $\left|\lambda_0\right\rangle$ and $\left|\lambda_{\pm}\right\rangle$ are the corresponding eigenfunctions of $-i\hat{S}^{(1)}(t)$. $\beta(t)$ can be written in terms of $\beta_0(t)=\mu_{10}\int_{t_0}^{t}dt'\mathcal{E}(t')\exp[i\omega_0t']$ and $\beta_1(t)=\mu_{21}\int_{t_0}^{t}dt'\mathcal{E}(t')\exp[i\omega_1t']$,  which are proportional to the Fourier transforms of the electric field at $\omega_0$ and $\omega_1$, respectively.
The corresponding wave function in the interaction picture  can be calculated by applying  $\hat{U}^{(1)}(t,t_0)$ onto  $|J_0M\rangle$. \\ \indent
For the molecule  starting from $J_0=0$ and $M=0$, the wave function of the system  can be given by
 \begin{eqnarray} \label{J0M0}
|\psi_{00}^{(1)}(t)\rangle&=&\frac{[\left|\beta_1(t)\right|^2+\left|\beta_0(t)\right|^2\cos\beta(t)]}{\beta^2(t)}|00\big\rangle\\ \nonumber
&&+\frac{i\beta^*_0(t)\sin\beta(t)}{\beta(t)}|10\rangle+\frac{\beta^*_0(t)\beta^*_1(t)}{\beta^2(t)}\left[\cos\beta(t)-1\right]|20\rangle,
\end{eqnarray}
 which can be interpreted as a rotational ladder-climbing mechanism, which has already identified in Ref. \cite{JCP2005} to produce molecular orientation from the rotational ground state $|00\rangle$. That is, a one-photon transition to $|10\rangle$ occurs at the frequency $\omega_0$, whereas the transition to $|20\rangle$ is a one-photon transition at the frequency $\omega_0$ followed by a one-photon transition at the frequency $\omega_1$, i.e., indirect (resonant) two-photon absorption via separate one-photon transitions.
 For the molecule  starting from $J_0=1$ and $M=0$, we can obtain  \begin{eqnarray}\label{J1M0}
|\psi_{10}^{(1)}(t)\rangle&=&\frac{i\beta_0(t)\sin\beta(t)}{\beta(t)}|00\big\rangle+\cos\beta(t)|10\rangle\\ \nonumber
&&+\frac{i\beta_1(t)\sin\beta(t)}{\beta(t)}|20\rangle.
 \end{eqnarray}
  We can see that the ratio of population transfer to $|00\rangle$ and $|20\rangle$ is determined by $\beta_0$ and $\beta_1$. Thus the molecules  absorb the photons at frequencies at $\omega_0$ and  $\omega_1$, resulting in two resonant one-photon transitions from $|10\rangle$ to $|00\rangle$ and $|20\rangle$, respectively.
For the molecule  starting from $J_0=1$ and $M=\pm1$, the transitions to the ground rotational state $J=0$ are forbidden, and thus the corresponding wave function reads
 \begin{eqnarray}\label{J1M1}
| \psi_{1M}^{(1)}(t)\rangle&=&\cos\beta(t)|1M\rangle+ie^{i\phi_c}\sin\beta(t)|2M\rangle.
 \end{eqnarray}
with $\beta(t)=|\beta_1(t)|$. It describes a resonant one-photon transition from $|11\rangle$ to $|21\rangle$ by absorbing the photon at frequency $\omega_1$. The details concerning the derivation of Eqs. (\ref{J0M0})-(\ref{J1M1})
 can be found  in Appendix \ref{A}.  \\ \indent
We now examine the OQR phenomena for the molecules starting from a pure rotational state. Since the shape of the single-cycle THz pulse depends on $\omega_c$ and $\mathcal{E}_0$, we perform simulations to show the dependence of the OQR amplitude on the two parameters. To consider the THz pulse with a comparable duration to the revival time, we vary $\omega_c$ from 0.072 to 0.108 THz, with a small deturning  $\delta_1=\omega_c-\omega_0$, and modulate $\mathcal{E}_0$  from $1.0\times10^5$ to $8.0\times10^6$ V/m.  Figures \ref{fig3} (a)-(c) plot the landscape of the OQR amplitude with respect to $\delta_1$  and $\mathcal{E}_0$ for the molecules initially in $J_0=0$ and $M=0$, and $J_0=1$ and $M=0, 1$, respectively. OQRs occur in all three cases,  and the OQR amplitude strongly depends on $\mathcal{E}_0$. We can  see from Eqs. (\ref{J0M0}) to (\ref{J1M1}) that $\beta_0(T)$ and $\beta_1(T)$  determine the probabilities of rotational states, requiring that the transition frequencies of the adjacent rotational states are within the frequency distribution of the THz pulse.  Figures \ref{fig3} (d)-(f) show comparisons of the exactly calculated $A_{OQR}$ versus $\mathcal{E}_0$ at  $\delta_1=0$ with that by expanding the unitary operator to the first- and third-order Magnus terms. For low field strengths, the OQR amplitudes within the three-state model by Eqs. (\ref{J0M0}-\ref{J1M1})  agree with the exact simulations. As the strength increases, the first-order descriptions start to deviate from the exact one in Figs. \ref{fig3} (d)-(f). It implies that the optical processes via high-order Magnus terms play roles in the strong field regime.   \\ \indent
\begin{figure}[!t]\centering
\resizebox{0.45\textwidth}{!}{%
  \includegraphics{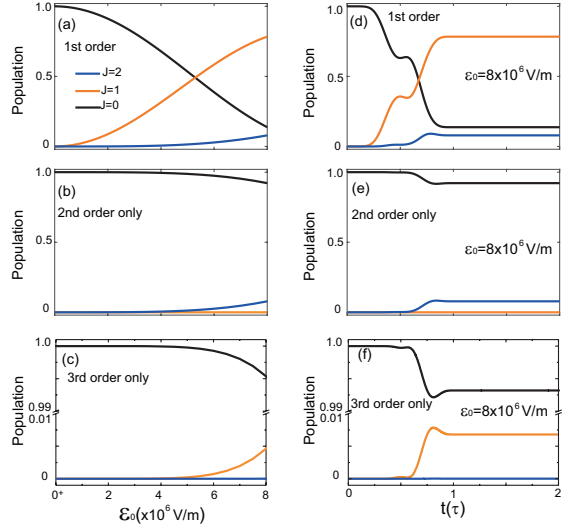}
} \caption{The final populations versus the field strength $\mathcal{E}_0$ for the molecules starting from $|00\rangle$ by only considering (a) the first- (b) second- and (c) third-order Magnus terms in the unitary time-evolution operator. (d)-(f) The corresponding time-dependent populations  at the field strength of $\mathcal{E}_0=8.0\times10^6$V/m. } \label{fig4}
\end{figure}
To  understand the effects of higher-order Magnus terms on OQRs,  we perform simulations by only considering one-order Magnus term in the time-dependent unitary operator.  As an example, we consider the system initially in the state $|00\rangle$, and the corresponding wave function of the system in the interaction picture can be written as $|\psi_{00}(t)\rangle^{(n)}_I=\exp\left[\hat{S}^{(n)}\left(t\right)\right]|00\rangle$. Figures \ref{fig4} (a)-(c) show the final populations versus $\mathcal{E}_0$ by only considering one order of the Magnus terms. The first-order Magnus term leads to quantum state transfer from $|00\rangle$ to $|10\rangle$ and then to  $|20\rangle$ in Fig. \ref{fig4} (a), in good agreement with the underlying processes by Eq. (\ref{J0M0}).
Thus, the transition to $|10\rangle$ is a one-photon transition at the frequency $\omega_0$, whereas
the transition to $|20\rangle$ is a one-photon transition at the frequency $\omega_0$ followed by
a one-photon transition at the frequency $\omega_1$. From Figs. \ref{fig4} (b) and (c), we can see that the optical transition processes via higher-order Magnus terms become visible  in the strong field strength regime. The second-order one leads to optical transition from the initial state $|00\rangle$ to the second rotational excited state $|20\rangle$ without population in the first rotational excited state $|10\rangle$. The third-order one induces the population transfer to $|10\rangle$ without population in  $|20\rangle$. These high-order Magnus terms lead to the underlying optical transitions going beyond the description of the first-order Magnus term, i.e., the  rotational ladder-climbing mechanism. \\ \indent
To further visualize the underlying optical transition processes, Figs. \ref{fig4} (d)-(f) show the time-dependent population transfer processes induced by only the first-,  second- or third-order Magnus term, respectively, at the field strength of $\mathcal{E}_0=8.0\times10^6$V/m for the molecules starting from $|00\rangle$. Different from the time-dependent population transfers in Fig. \ref{fig4} (d), we can see that the second-order term does not induce any population transfer to $|10\rangle$ during the whole interaction of the THz pulse in Fig. \ref{fig4} (e). It  indicates that the second-order term opens the  transition pathways from $|00\rangle$ to $|20\rangle$ with simultaneous two-photon absorption, which can be viewed as direct two-photon transitions by the interaction of the photons with the molecule. The optical transition from $|00\rangle$ to $|10\rangle$ via the third-order term occurs at the strong-field regime in Fig. \ref{fig4} (f) and it does not induce any further transition from $|10\rangle$ to $|20\rangle$, which is also different from the optical process  induced by the first-order term. It implies that the third-order Magnus term opens transition pathways from $|00\rangle$ to $|10\rangle$ with simultaneous three-photon absorption. As a result, these direct  multiphoton processes can be induced via the higher-order Magnus terms, which will interfere  with the optical processes via the first-order term, leading to the OQR enhancement in Fig. \ref{fig3} (d) \\ \indent
\begin{figure}[!t]\centering
\resizebox{0.45\textwidth}{!}{%
  \includegraphics{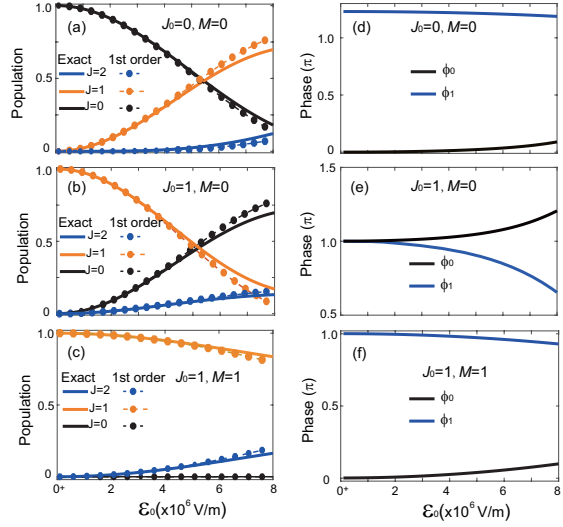}
} \caption{The final populations and the relative phases with respect to the field strength $\mathcal{E}_0$ at $\delta_1=0$. (a)-(c) The final populations versus $\mathcal{E}_0$ for molecules starting from $|00\rangle$, $|10\rangle$ and $|11\rangle$, respectively, which are compared with that by using the first-order Magnus term.  (d)-(f)  The corresponding relative phases versus $\mathcal{E}_0$. Note that all phases are wrapped to the range [0, $2\pi$].} \label{fig5}
\end{figure}
Figure  \ref{fig5} shows the final rotational populations and the corresponding relative phases $\phi_0$ and $\phi_1$.  The final populations in Figs. \ref{fig5} (a)-(c) follow the first-order descriptions very well for low strengths. As the field strength increases, the differences between the exact and first-order simulations become visible. Thus, the higher-order Magnus terms  can activate nonresonant multiphoton transitions (from the initial state to a given final state by absorbing  multiple photons
simultaneously without involving the intermediate states. For the  superposition consisting of three rotational states in Figs. {\ref{fig5}} (a) and (b), the corresponding degree of orientation reads $\langle\cos\theta\rangle(t)=2/\sqrt{3}|c_{10}||c_{00}|\cos(\omega_0t-\phi_0)+4/\sqrt{15}|c_{20}||c_{10}|\cos(\omega_1t-\phi_1)$. Thus, the relative phases between neighboring rotational states affect the maximal degree of orientation via quantum interference between pairs of rotational states \cite{J2}. From Eqs. (\ref{J0M0}-\ref{J1M1}), we can see that the phase of each rotational state does not change with respect to $\mathcal{E}_0$.  The noticeable changes in the relative phases (see Figs. \ref{fig5} (d) and (e)) can be attributed to the optical processes via higher-order Magnus terms, capable of enhancing the OQR  amplitude over the level by Eqs.  (\ref{J0M0}-\ref{J1M1}). For the molecule starting from $J_0=1$ and $M=\pm1$ in Fig. \ref{fig5} (c), the superposition of rotational states $|11\rangle$ and $|21\rangle$ reduces the expression for the degree of orientation to $\langle\cos\theta\rangle(t)=2/\sqrt{5}|c_{21}||c_{11}|\cos(\omega_1t-\phi_1)$, leading to the OQR amplitude $A_{OQR}=4/\sqrt{5}|c_{21}||c_{11}|$, which  is independent of $\phi_1$ and reaches its maximum at  $2/\sqrt{5}$ (i.e., 0.89) with equal weights of $c_{11}$ and $c_{21}$. For such a two-state system,  the  optical processes via the higher-order  Magnus terms suppress population transfer from $|11\rangle$ to $|21\rangle$ and decreases the value of $A_{OQR}$ below the level of the first-order Magnus description in Fig. \ref{fig3} (f).\\ \indent
 We finally discuss the feasibility of performing the present scheme in experiments. For restricting the problem into the three-state model, the field strengths used in the simulations are below the limit of the reported  THz pulses \cite{SCTP}. If we further increase the field strength, the optical processes may become more complex, e.g., by involving higher rotational states of $J>2$ into the wave packets.  To that end, we examine the molecule initially in $|00\rangle$ by using the experimentally reported 0.1 THz  pulses. Figure \ref{fig6} shows the dependence of the final populations and the corresponding local maximum of $|\langle\cos\theta\rangle|$ on $\mathcal{E}_0$. There are visible populations in the state of $J=3$ for $\mathcal{E}_0>1.0\times10^7$ V/m. Interestingly, however, the degree of orientation reaches a local maximum of $|\langle\cos\theta\rangle|_{max}=0.78$ at $\mathcal{E}_0=0.91\times10^7$, resulting in negligible population in $|30\rangle$. \\ \indent
 \begin{figure}[!t]\centering
\resizebox{0.48\textwidth}{!}{%
  \includegraphics{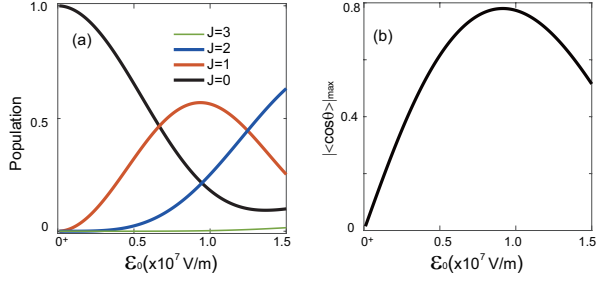}
} \caption {OQRs induced by an experimentally accessible single-cycle strong-field 0.1 THz source. (a) The final populations in the rotational states versus the field strength $\mathcal{E}_0$ for the molecules starting from the ground rotational state $|00\rangle$, (b) the corresponding maximal values of $|\langle\cos\theta\rangle|$. } \label{fig6}
\end{figure}
 Based on the above analysis, the realization of the three-state OQR is expected for molecules at ultracold temperatures. Experimentally a two-state model for  OQR has been demonstrated for absolute-ground-state-selected OCS molecules by the combination of a 485-ps-long nonresonant laser pulse and a weak static electric field \cite{ts},  obtaining a  value of $\approx0.577$ for the degree of orientation, i.e., the theoretical maximum $\sqrt{1/3}$ for the two-state model starting from $|00\rangle$.  The present three-state model without the use of the static electric field can reach this theoretical maximum at the field strength of $\mathcal{E}_0=0.46\times10^7$ V/m (see Fig. \ref{fig6} (b)) with a small amount of population in the state of $J=2$ in Fig. \ref{fig6} (a).
 Note that the three-state scheme will reduce to a two-state model by further increasing the duration of the THz pulses while keeping the resonant excitation condition, e.g., by using a multi-cycle THz pulse.
 For practical applications, the wave packet consisting of the lowest (two, or three) rotational states in a lower-dimensional subspace  is beneficial to obtaining a long duration of OQR with its amplitude above a given threshold \cite{sugny2}. \\ \indent
 The present method can be generally applied to other molecules by matching the central frequency and the peak field strength of the single-cycle THz pulse. A fundamentally important question remains open whether the OQR amplitude within the three-state model can be optimized by tailoring the THz pulse with a constraint of zero pulse area  \cite{shupra,Rabitz2013,jcp:OCT}. By fixing the power spectrum of the THz pulse, a spectral phase-only optimization \cite{QOCT2,shu6,IEEEDong,IEEEshu} may enhance the OQR amplitude by modulating the relative phases $\phi_J$ between pairs of neighboring rotational states.\\ \indent
In summary, we theoretically examined OQRs in molecules by using a zero-area-single-cycle THz pulse with the comparable duration to the revival time and performed the simulations for the linear polar molecule HCN with experimentally available pulse parameters. A large OQR occurs even at finite temperatures without the additional use of an intense nonresonant pulse or a static electric field.  We analyzed the underlying physics within the three-state model.  By performing the Magnus expansion of the time-evolution operator, it reveals that the physical processes via higher-order Magnus terms can enhance the OQR amplitude over the level by the first-order  Magnus term. We also examined the experimental feasibility of the present scheme for generating a three-state OQR. This work provides an fundamentally important insight into the THz-laser-induced  field-free molecular orientation, which has a wide variety of applications ranging from molecular-phase modulators, ultrafast X-ray diffraction, and ultrashort pulse compression to chemical reactivity, nanoscale design, and high harmonic generation.
 \begin{acknowledgements}
 All authors are  grateful to their family for their great support so that they can work efficiently at home during the COVID-19 outbreak.  This work was supported by the National Natural Science Foundations of China (NSFC) under Grant No. 61973317. Y. G. is partially supported by 
the Opening Project of Key Laboratory of Low Dimensional
Quantum Structures and Quantum Control of the Ministry of
Education under Grant No. QSQC1905.
\end{acknowledgements}

\appendix
\section{}\label{A}
We consider a model consisting of three states $|00\rangle$, $|1M\rangle$, and $|2M\rangle$ with energies $E_0$, $E_1$ and $E_2$, which is driven by  a linearly polarized time-dependent laser pulse $\mathcal{E}(t)$ via the interaction with the electric dipole moment $\mu$ with elements $\mu_{01}=\mu_{10}$ and $\mu_{12}=\mu_{21}$.
The corresponding time-dependent Hamiltonian of the system reads
\begin{align}
\hat{H}(t) =\left(\begin{array}{ccc}
E_{0} & 0 & 0\\
0 & E_{1} & 0\\
0 & 0 & E_{2}
\end{array}\right)-\left(\begin{array}{ccc}
0 & \mu_{10} & 0\\
\mu_{10} & 0 & \mu_{21}\\
0 & \mu_{21} & 0
\end{array}\right)\mathcal{E}(t).
\end{align}\\ \indent
We write the Hamiltonian in the interaction picture without using the rotating wave approximation,
\begin{equation}
\hat{H}_{I}(t)=-\left(\begin{array}{ccc}
0 & \mu_{10}\mathcal{E}(t)e^{-i\omega_{0}t} & 0\\
\mu_{10}\mathcal{E}(t)e^{i\omega_{0}t} & 0 & \mu_{21}\mathcal{E}(t)e^{-i\omega_{1}t}\\
0 & \mu_{21}\mathcal{E}(t)e^{i\omega_{1}t} & 0
\end{array}\right),
\end{equation}
with $\omega_{0}=(E_{1}-E_{0})$ and $\omega_{1}=(E_{2}-E_{1})$. The time-dependent wave function of the system starting from a given initial state $|i\rangle$ can be given by $|\psi(t)\rangle_I=\hat{U}(t,t_0)|i\rangle$ with a unitary operator $\hat{U}(t, t_0)$ and $\hat{U}(t_0,t_0)=\mathbb{I}$.\\ \indent
To obtain an analytical solution of $|\psi(t)\rangle_I$, we expand the unitary operator $\hat{U}(t, t_0)$ by using Magnus expansion \cite{pr:470:151}
\begin{equation}
\hat{U}(t,t_{0})=\exp\Bigg[\sum_{n=1}^{\infty}\hat{S}^{(n)}(t)\Bigg]
\end{equation}
where the first three leading terms can be given by means of the Baker-Campbell-Hausdorff formula as $\hat{S}^{(1)}(t)=-i\int_{t_0}^tdt_1\hat{H}_I(t_1)$, $\hat{S}^{(2)}(t)=(-i)^2/2\int_0^t dt_1\int_0^{t_1}dt_2[\hat{H}_I(t_1),\hat{H}_I(t_2)]$, and $\hat{S}^{(3)}(t)=(-i)^3/6\int_0^t dt_1\int_0^{t_1}dt_2\int_0^{t_2}dt_3[\hat{H}_I(t_1),[\hat{H}_I(t_2), \hat{H}_I(t_3)]]$.\\ \indent
We now consider the case by solely involving
the first-order term in the Magnus expansion, which can be defined by $\hat{S}^{\left(1\right)}\left(t\right)=iA\left(t\right)$
with
\begin{align}
A\left(t\right) & =-\int_{t_{0}}^{t}H_{I}\left(t'\right)dt'\nonumber \\
 & =\left(\begin{array}{ccc}
0 & \beta_{0}^{*}(t) & 0\\
\beta_{0}(t) & 0 & \beta_{1}^{*}(t)\\
0 & \beta_{1}(t) & 0
\end{array}\right)
\end{align}
where $\beta_{0}(t)=\mu_{10}\int_{t_{0}}^{t}dt'\mathcal{E}(t')e^{i\omega_{0}t'}$ and $\beta_{1}(t)=\mu_{21}\int_{t_{0}}^{t}dt'\mathcal{E}(t')e^{i\omega_{1}t'}$. \\ \indent
 By diagonalizing the matrix $\hat{S}^{(1)}(t)$, the unitary operator  to
the first-order term $\hat{S}^{(1)}(t)$ reads
\begin{eqnarray}
\hat{U}^{(1)}(t,t_{0}) & = &\exp(iA(t))\nonumber \\
&=&\sum_{p=-,0, +}\exp(i\lambda_{p}(t))|\lambda_{p}\rangle\langle\lambda_{p}|
\end{eqnarray}
where $\lambda_{0}(t)=0,\lambda_{-}(t)=-\beta(t)$, and $\lambda_{+}(t)=\beta(t)$
are the eigenvalues of $S^{(1)}(t)$, and the corresponding eigenstates
are
\begin{align}
|\lambda_{0}\rangle & =\frac{|\beta_{0}(t)|}{\beta(t)}\left(-\frac{\beta_{1}(t)}{\beta_{0}^{*}(t)}|00\rangle+|2M\rangle\right),\\
|\lambda_{-}\rangle & =\frac{1}{\sqrt{2}}\frac{|\beta_{1}(t)|}{\beta(t)}\left(\frac{\beta_{0}(t)}{\beta_{1}^{*}(t)}|00\rangle-\frac{\beta(t)}{\beta_{1}^{*}(t)}|1M\rangle+|2M\rangle\right),\\
|\lambda_{+}\rangle & =\frac{1}{\sqrt{2}}\frac{|\beta_{1}(t)|}{\beta(t)}\left(\frac{\beta_{0}(t)}{\beta_{1}^{*}(t)}|00\rangle+\frac{\beta(t)}{\beta_{1}^{*}(t)}|1M\rangle+|2M\rangle\right),
\end{align}
with $\beta(t)=\sqrt{|\beta_{0}(t)|^{2}+|\beta_{1}(t)|^{2}}$.
The corresponding wave functions in term of the first-order Magnus expansion  can be calculated by applying  $\hat{U}^{(1)}(t,t_0)$ onto  $|J_0M\rangle$, i. e., $|\psi_{J_0M}^{(1)}(t)\rangle=\hat{U}^{(1)}(t, t_0)|J_0M\rangle$, which will lead to Eqs. (\ref{J0M0})-(\ref{J1M1}).

\end{document}